\documentclass[twocolumn]{aa}
\usepackage{epsfig}
\usepackage{graphicx}
\usepackage[latin1]{inputenc}
\def\simlt{\ \raise -2.truept\hbox{\rlap{\hbox{$\sim$}}\raise5.truept   %
\hbox{$<$}\ }}
\def\simgt{\ \raise -2.truept\hbox{\rlap{\hbox{$\sim$}}\raise5.truept   %
\hbox{$>$}\ }}                                                          %

\def\be{\begin{equation}}
\def\ee{\end{equation}}
\def\newline{\hfil\break}

\def\la{\mathrel{\hbox{\rlap{\hbox{\lower4pt\hbox{$\sim$}}}\hbox{$<$}}}}
\def\ga{\mathrel{\hbox{\rlap{\hbox{\lower4pt\hbox{$\sim$}}}\hbox{$>$}}}}

\begin{document}
\title{The energetics of giant radio galaxy lobes from inverse Compton scattering observations}
   \author{S. Colafrancesco\inst{1,2} and P. Marchegiani\inst{3}}
   \offprints{S. Colafrancesco}
\institute{
              School of Physics, University of the Witwatersrand,
              Johannesburg Wits 2050, South Africa.
              Email: sergio.colafrancesco@wits.ac.za
 \and
              INAF - Osservatorio Astronomico di Roma
              via Frascati 33, I-00040 Monteporzio, Italy.
              Email: sergio.colafrancesco@oa-roma.inaf.it
 \and
              Dipartimento di Fisica, Universit\`a di Roma La Sapienza, P.le A. Moro 2, Roma, Italy
              Email: paolo.marchegiani@oa-roma.inaf.it
             }
\date{Received  / Accepted  }
\authorrunning {S. Colafrancesco and P. Marchegiani}
\titlerunning {Energetics of GRG lobes from ICS emission}
\abstract
  %{} leave it empty if necessary
   {Giant radio galaxy (GRG) lobes are excellent laboratories to study the
   evolution of the particle and magnetic field energetics, and the
   past activity of radiogalaxy jets, as indicated by recent results of
   X-ray observations with Suzaku. However, these results are based on
   assumptions of the shape and extension of the GRG lobe electron spectrum.}
   {%Aims.
   We re-examine the energetics of GRG lobes as derived by inverse
   Compton scattering of CMB photons (ICS-CMB) by relativistic electrons in RG lobes to assess
   the realistic physical conditions of RG lobes, their energetics and their radiation
   regime. We consider the steep-spectrum GRG DA 240 recently observed by Suzaku as a
   reference case and we also discuss other RG lobes observed with Chandra and XMM.}
  {%Method.
  We model the spectral energy distribution (SED) of the GRG DA 240 East lobe to get constraint on the shape and on the
  extension of the electron spectrum in the GRG lobe by using multi-frequency
  information from radio to gamma-rays. We use radio and X-ray data to constrain
  the shape and normalization of the electron spectrum
  and we then calculate the SZ effect expected in GRG lobes that is sensitive to the total
  electron energy density. }
   {%Results.
    We show that the electron energy density $U_e$ derived form X-ray observations
    can yield only a rough lower limit to its actual value and that most of the
    estimates of $U_e$ based on X-ray measurements have to be increased even by a large factor
    by considering realistic estimates of the lower electron momentum $p_1$. This brings
    RG lobes away from the equipartition condition towards a particle-dominated and
    Compton power dominance regime.
    We propose to use the distribution of RG lobes in the $U_e/U_B$ vs. $U_e/U_{CMB}$
    plane as a further divide between different physical regimes of particle and field
    dominance, and radiation mechanism dominance in RG lobes.}
  {%Conclusions.
  We conclude that the SZ effect produced by ICS-CMB mechanism
  observable in RG lobes provides reliable
  estimate of $p_1$ and $U_e$ and is the best tool to determine
  the total energy density of RG lobes and to assess their physical regime.
  This observational tool is at hand with the sensitive high-frequency radio and mm
  experiments.}

 \maketitle
%----------------------------------------------------

\section{Introduction}
 \label{sec.intro}

Giant radio galaxy (GRG) lobes are considered the final stage of
the evolution of radiogalaxy (RG) jets and are likely the
evolutionary connection with the inverse Compton ghosts of GRGs.
The details of such evolutionary process are, however, not yet
well understood. In this context, GRG lobes are excellent
laboratories to study the evolution of the particle and magnetic
field energy density, and the past activity of RG jets.

Determining the total particle and magnetic energy density in GRG
lobes is a difficult task because of the lack of precise
indicators of the overall particle energy spectrum (especially in
the low-energy part, where most of the energy comes from for steep
spectrum sources) and of the distribution and power spectrum of
the magnetic field.\\
The electron and magnetic energies stored in GRG lobes are usually
measured using the diffuse radio synchrotron emission and the
inverse Compton scattering (ICS) X-ray emission where these X-rays
are produced by cosmic microwave background (CMB) photons
up-scattered by the ICS with the relativistic electrons (hereafter
ICS-CMB) injected in the GRG lobe.\\
The results of the ICS-CMB X-ray emission from RG lobes obtained
with ROSAT (e.g. Feigelson et al. 1995), ASCA (e.g., Kaneda et al.
1995; Tashiro et al. 1998, 2001), Chandra (e.g., Isobe et al.
2002; Croston et al. 2005; Yaji et al. 2010), XMM-Newton (e.g.,
Isobe et al. 2005; Isobe et al. 2006) and Suzaku (Isobe et al.
2009, 2011a,b) indicate a dominance of the electron energy density
$U_e$ over the magnetic one $U_B$ by typically an order of
magnitude in many RG lobes (see Croston et al. 2005; Isobe et al.
2009, 2011a,b), even though several lobes are also found in almost
equipartition condition $U_e = U_B$ (see, e.g., discussion in
Isobe et al. 2011a).

The Suzaku satellite recently measured ICS-CMB X-ray emission from
a few GRG lobes (e.g. 3C326, Isobe et al. 2009; DA240, Isobe et
al. 2011a; 3C35, Isobe et al. 2011b) and these results seem to
indicate that GRG lobes have very low particle and magnetic field
energy density that would set these objects at the lower end of
the $U_B-U_e$ correlation. These results strengthen the idea that
the dominance of the ICS-CMB radiative losses over the synchrotron
one is a common properties in the lobes of GRGs (see, e.g.,
Ishwara-Chandra \& Saikia 1999), and that RG lobes developing from
a size $D \simlt 100$ kpc to a size $D \sim $ Mpc induce an
evolution of their energetics from the electron dominance regime
(mainly in the jet) to the equipartition regime (mainly in the
lobe), following the adiabatic evolution $U_e \propto D^{-2}$, in
addition to a significant decrease in both $U_e$ and $U_B$ (see
discussion in Isobe et al. 2011a).

The specific entity of these results depend, however, on a
detailed knowledge of the electron spectrum over the whole range
of electron energies, an issue that is not easily obtainable from
X-ray ICS-CMB observations which are sensitive to electron Lorentz
factors $\gamma \simgt 10^3$  (i.e. only in the high-energy branch
of the electron spectrum) while most of the CMB photon
Comptonization is produced in the low-energy part of the electron
spectrum (see, e.g., Croston et al. 2005, Colafrancesco 2008).
This is especially so for GRG lobes with steep energy spectra.

We re-examine here the derivation of the GRG lobe energetic and
B-field estimates based on a detailed modeling of the
multi-frequency ICS-CMB spectra and of their associated
synchrotron spectra.
To provide a specific and quantitative discussion, we discuss the
reference case of the GRG DA 240 observed recently with Suzaku
(Isobe et al. 2011a) that is the object with the steepest energy
spectrum in its radio lobes among those observed with Suzaku. The
GRG DA 240 is also one of the steepest spectra objects with
respect to the RG sample discussed in Croston et al. (2005).\\
Steep energy spectra make the determination of $U_e$ more
uncertain by using only X-ray observations, and the strong
sensitivity to the value of the minimum electron energy requires a
more refined analysis.
In this respect, we explore the parameter space region allowed by
the data in order to study the amount of variation on the electron
energy density $U_e$.
We show in Sect.2 that the X-ray estimates of $U_e$ provide only a
rough lower limit for $U_e$ and of the minimum energy
$\gamma_{min}$ of the electron population, and we discuss the
consequences of such assumptions on the RG lobe energy and
radiation regime and on their multi-frequency spectral energy
distributions (SEDs).
We then show in Sect.3 that observations of the SZ effect produced
in GRG lobes (as originally proposed by Colafrancesco 2008) are
the only ones able to provide the missing information on the total
electron energy density $U_e$ and also provides reliable estimates
of the quantity $U_e/U_B$ and of the structure and evolution of
GRG lobes. We discuss in Sect.4 how similar arguments apply to
other RG lobes studied in the X-ray energy band, and we discuss
the various physical regimes in which RG lobes are found based on
their distribution in the $U_e/U_B$ vs. $U_e/U_{CMB}$ plane. We
summarize our results and draw our conclusions in Sect.5.

Throughout the paper, we use a flat, vacuum--dominated
cosmological model with $\Omega_m = 0.27$, $\Omega_{\Lambda} =
0.73$ and $H_0 = 71$ km s$^{-1}$ Mpc$^{-1}$.

\section{A tale of a GRG: DA 240}

The GRG DA 240 at $z=0.03561$ (Rines et al. 2000) has a radius of
the East lobe, in which the ICS-CMB X-ray emission has been
measured by Suzaku, of $\sim7$ arcmin (Isobe et al. 2011a), which
correspond to a linear size of $294$ kpc at the GRG redshift. This
is a prototype of GRG with extended lobes that is useful to test
our analysis.\\
The radio synchrotron spectral index in the East lobe is
$\alpha_r=0.95\pm0.01$ (Mack et al. 1997) in the range 326 to 609
MHz, and flattens at higher frequency reaching a value $\approx
0.74$. The radio spectrum measured between 4.8 and 10.6 GHz
appears to be even flatter with  $\alpha_r \approx 0.58$, since
the hot spot emission is likely dominant in this high frequency
range.\\
The X-ray ICS-CMB emission measured by Suzaku has a spectral index
$\alpha_X=0.92^{+0.13, +0.04}_{-0.17, -0.06}$  compatible with the
radio spectral index $\alpha_r$ measured at low-$\nu$ between 326
and 609 MHz, but certainly steeper than the radio spectral index
measured between 2.7 and 10.5 GHz (see Mack et al. 1997).\\
Under the assumption $\alpha_X = \alpha_r$, Isobe et al. (2011a)
derived the X-ray ICS-CMB flux $F_{1keV} = 51.5 \pm
3.9^{+6.2}_{-5.4}$ nJy (with a best-fit reduced $\chi^2$ smaller
w.r.t. the fit leaving $\alpha_X$ as a free parameter), an overall
B-field value inside the lobe of $B =0.87\pm 0.03^{+0.05}_{-0.06}$
$\mu$G corresponding to $U_B \approx (3.0 \pm 0.2\pm0.4)\cdot
10^{-14}$ erg cm$^{-3}$, and a value $U_e = (3.4^{+0.3;
+0.5}_{-0.2; -0.4}) \cdot 10^{-14}$ erg cm$^{-3}$ integrated over
the range of electron energies $\gamma = 10^3-10^5$. Here $\gamma
= E/m_e c^2$.
The previous analysis done using $\gamma_{min}=10^3$ yields
$U_e/U_B \approx 1.1$, $U_B/U_{CMB} \sim 0.06$ (where $U_{CMB}
\approx 4.8 \cdot 10^{-13}$ erg cm$^{-3}$ at the redshift of DA
240) and $U_e/U_{CMB} \approx 0.07$.

We stress that the previous values of the particle energy density
$U_e$ depends strongly on the assumed electron spectrum at low
energy (where most of the ICS-CMB originates) and specifically
from the spectral index $\alpha$ and from the minimum momentum
$p_1$ (or equivalently $\gamma_{min}$) of the electron
population.\\
We notice that the previous values of $U_e$ and $U_B$ have been
obtained by Isobe et al. (2011a) by deriving the electron density
and the B-field intensity using the approximate analytical
formulae given in Harris \& Grindlay (1979) that are based on
various assumptions: the monochromatic approximation for the ICS
emission according to which an electron with energy $E$ emits by
ICS at a single frequency derived by assuming a monochromatic CMB
photon spectrum; the monochromatic approximation for the
synchrotron emission according to which an electron with energy
$E$ in a magnetic field with intensity $B$ emits a a single
frequency.\\
In the following we make use of the complete formulae for the
synchrotron and ICS emission mechanisms (see, e.g., Schlickeiser
2002), that yield the correct results of the same order of
magnitude of the approximate relations, but with some numerical
difference in some specific case with respect to the approximated
calculations used in Harris \& Grindlay (1979).

To address the issue of a precise determination of $U_e$, we use a
power-law electron spectrum to model the low-$\nu$ part of the
radio spectrum
\begin{equation}
N(p)=k_0 p^{-\alpha}   \;\;\;\; p\geq p_1,
\end{equation}
where $p=\beta\gamma$.
Given this electron spectrum, the total energy density of the
electron population in the RG lobe can be directly obtained (see
Colafrancesco 2008, Colafrancesco et al. 2011, En\ss lin \& Kaiser
2000, Colafrancesco et al. 2003) as
\begin{equation}
U_e=\int_{p_1}^\infty dp N(p) (\sqrt{1+p^2}-1) m_e c^2 \; .
\label{eq.ene.ele}
\end{equation}

We consider, for the sake of generality, values of the minimum
electron momentum in the range $p_1=1-10^3$. A value $p_1 \approx
10^3$ corresponds to the lower energy observable in the X-ray
band, of order of $\sim 1$ keV (see Isobe et al. 2011a), while
values $p_1\sim 10^2-10^3$ are derived from observations of
hotspots (see e.g., Carilli et al. 1991), and even lower values
$p_1\sim 1-10$ are expected due to the effects of adiabatic
expansion of the RG lobe (see, e.g., discussion in Croston et al.
2005). Exploring a wide range of possible values for $p_1$ is
crucial for our discussion since its specific value plays a
crucial role in the derivation of the actual value of $U_e$ (see
eq.\ref{eq.ene.ele}). We will discuss a new observational strategy
aimed at the derivation of $p_1$ in Sect.3 below.

The measured value of $\alpha_r$ at low frequencies corresponds to
a value of the electrons spectral index $\alpha=2.9$ (here
$\alpha= 2 \alpha_r +1$) for the spectrum at low energies.

An electron spectrum with a break $p_b$ at low energy, with
$\alpha=\alpha_1$ for $p < p_b$ and $\alpha=\alpha_2$ for $p>
p_b$, as produced by electron energy losses, effectively behaves
for our purposes as the single power-law spectrum but with the
difference that here the crucial quantity (both for the
determination of $U_e$ and for the spectral properties of the SZE)
is the value of the break $p_{b}$ instead of $p_1$.

\begin{figure}[ht!]
\begin{center}
 \epsfig{file=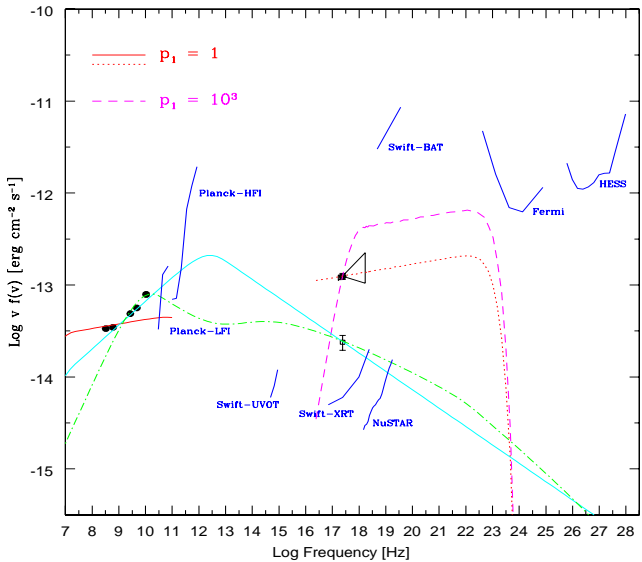,height=8.cm,width=9.cm,angle=0.0}
\end{center}
 \caption{\footnotesize{The SED of the East lobe of DA 240 calculated
 with a power-law spectrum recovering the low$-\nu$ radio data for
 $p_1=1$ and $B_{\mu} = 1.5$ (red solid for the radio spectrum and red dotted
 for the high-$\nu$ spectrum) and $p_1=10^3$ $B_{\mu}=0.87$ (magenta dashed
 for the high-$\nu$ spectrum).
 The electron spectrum has a cutoff at $p_{cut} = 5 \cdot 10^5$ in order to produce an
 ICS-CMB gamma-ray emission below the Fermi-LAT 5-yr sensitivity limit that is
 exceeded for the case $p_1=10^3$ with a power-law spectrum without high-$\nu$ cutoff.
 A SSC SED model for the hot spot that fits the X-ray flux at 1 keV (open square point plus errorbars)
 with the synchrotron branch (cyan solid) or with the Compton branch (green dashed)
 is also shown for comparison.
 Various instrument sensitivities for point-like sources are shown for
 comparison. Instrument sensitivities are taken from the ASI-ASDC
 SED builder tool at http://tools.asdc.asi.it/
 }}
 \label{fig.sedpl2}
\end{figure}

We notice that the radio spectrum of the East lobe of DA 240 is
dominated at high frequencies by the hot spot. For this reason, we
decide to describe the hot spot emission with a single
self-synchrotron Compton (SSC) model that reproduces both the
high-$\nu$ radio data and the X-ray Suzaku observation at 1 keV.
The X-ray spectral shape around 1 keV of the hotspot is rather
uncertain with $\alpha_X$ in the range 0.72--1.27 (Isobe et al.
2011a, Evans et al. 2008) allowing decreasing or increasing SED
shape in the $\nu f(\nu) - \nu$ plot of Fig. \ref{fig.sedpl2}. We
calculated the SED of the hotspot with a single zone SSC model
with parameters describing the X-ray flux by synchrotron (cyan
solid) or by ICS (dashed green), since the available data do not
allow to distinguish between the two alternatives. In both cases
the hotspot is not dominating the overall SED of the East lobe of
DA 240 in the X-ray and gamma-ray frequency range, thus confirming
the idea that most of the gamma-ray emission associated to GRG
lobes comes from ICS-CMB emission from the diffuse electron
population in the radio lobes (see, e.g., the recent results on
the Fermi-LAT observations of the CenA lobes, Abdo et al. 2010).

We describe the total SED of the lobe as the sum of a power-law
spectrum SED model  for the diffuse electron population and a SSC
model for the hot spot.

We discuss in the following the multi-frequency SED of the diffuse
electron population of the East lobe of DA 240 to better constrain
the shape of the electron spectrum in this radio lobe.
We assume that electrons and B-field are distributed uniformly
within the spherical volume of the GRG lobe with a radius equal to
that of the projected area of the lobe.
We normalize the ICS-CMB emission expected with the previous
spectrum with $p_1=1$ to the X-ray flux observed by Suzaku at 1
keV (see Isobe et al. 2011a) and we obtain $k_0=1.02\times10^{-5}$
cm$^{-3}$.
This spectrum fits the synchrotron radio spectrum observed in the
range 326 - 609 MHz (Mack et al. 1997) with a value of the
magnetic field $B_{\mu}=1.5$, where $B_{\mu}$ is the magnetic
field in $\mu$G units.
The previous values of the normalization $k_0$ and of the B-field
are independent of $p_1$ for values lower than the value of $p$ at
which electrons emit by ICS at the observed energy $E_X$: for an
observed energy $E_X=1$ keV one obtains a value of $p\sim685$ by
using the approximate relations that an electron with energy $E_e$
emits ICS-CMB at the energy
\begin{equation}
E_X\sim8 \left(\frac{E_e}{\mbox{GeV}}\right)^2 \mbox{ keV}.
\label{eq.ICS.approx}
\end{equation}
and, correspondingly, synchrotron at a frequency
\begin{equation}
\nu\sim16 \left(\frac{E_e}{\mbox{GeV}}\right)^2
\frac{B}{\mu\mbox{G}} \mbox{ MHz} \; .
\end{equation}
This explains why the value of $k_0$ is independent of $p_1$ in
the range $p \approx 1-10^2$, while the situation changes for $p_1
\sim 10^3$. The difference in the values of $k_0$ and $B$ that we
obtain here for this last case and those derived by Isobe et al.
(2011a) is hence due to the different calculations performed (see
also discussion above).

The same power-law spectrum (with $\alpha = 2.9$) but with a
higher $p_1=10^3$ fits the Suzaku X-ray flux at 1 keV with a
normalization of $k_0=3.04\times10^{-5}$ cm$^{-3}$ and  the
low-$\nu$ radio spectrum with a magnetic field of $B_{\mu} = 0.87$
(in agreement with the result of Isobe et al. 2011a).

Fig.\ref{fig.sedpl2} shows the multi-frequency SED of the East
lobe of DA 240 for the two extreme values $p_1=1$ and $p_1=10^3$
to encompass the possible realistic values of this relevant
parameter.
The SED of DA 240 shows that the X-ray ICS-CMB emission spectrum
measured by Suzaku is consistent only with the low frequency radio
data. In addition, the X-ray ICS-CMB SED for $p_1=10^3$ has a
shape that is not consistent (within the uncertainties) with the
slope measured by Suzaku under the assumption of using the same
radio spectral index. This is because the minimum value $p_1=10^3$
reflects in a cutoff in the X-ray spectrum as shown in
Fig.\ref{fig.sedpl2} (see also discussion above).\\
We also find that the extrapolation at high frequencies of the
ICS-CMB power-law spectrum for $p_1=10^3$ is inconsistent with the
Fermi-LAT sensitivity limit, and therefore a high-E cutoff must be
introduced in the electron spectrum for this case. We adopt a
cutoff at $p_{cut} = 5 \cdot 10^5$ in order to be consistent with
the Fermi-LAT sensitivity limit, and the presence of this spectral
cutoff modifies both the ICS-CMB emission SED and the shape of the
radio SED. We adopt, for consistency, the same cutoff in the
electron spectrum also for the cases with $p_1=1$.

To summarize, the SED analysis of the ICS-CMB emission constrains
the electron spectrum to have i) a high-E cutoff at $E_{e,cut}
\sim 255$ GeV, ii) a spectral shape at low energies of $\alpha
\approx 2.9$, iii) a minimum momentum  $p_1$ certainly lower than
$10^3$ but not yet determined by the available X-ray data.

It is important to stress that X-ray ICS-CMB data provide the
normalization of the electron spectrum at $E_e=0.35 \mbox{ GeV}
(E_X/\mbox{keV})^{1/2}$ and the spectral index of the electron
spectrum in the energy range corresponding to the instrument
sensitivity ($\Delta E_e = 0.29-0.93$ GeV for the Suzaku window
$E_X=0.7-7$ keV), but these measurements are not sufficient to
determine the total energy density of the electron population that
depends strongly on the value of $p_1$, which cannot be determined
by typical X-ray ICS-CMB observations, as discussed in
Colafrancesco (2008) and Colafrancesco et al. (2011).
As a consequence, all the considerations on the ratio $U_e/U_B$,
the possible equipartition regime, and particle domination regimes
are misleading and need to be revised on the basis of a more
realistic estimate of the total electron energy density. This is
possible by using another observational feature of the ICS-CMB
emission, i.e. the SZ effect (SZE) from the RG lobe (see
Colafrancesco 2008), as we discuss in the next section.

\section{The SZE from the East lobe of DA 240}

It has been shown (Colafrancesco 2008, see also Colafrancesco et
al. 2011) that the same ICS-CMB mechanism that accounts for the
X-ray emission from the lobes of GRGs also provides a SZE from GRG
lobes and that its amplitude and spectral characteristics depend
on the total electronic energy density of the GRG lobe. The CMB
intensity change $\Delta I_{\rm lobe}(x)$ due to the SZE in the
lobe writes as
\begin{equation}
\Delta I_{\rm lobe}(x)=2\frac{(k_{\rm B} T_{CMB})^3}{(hc)^2}y_{\rm
lobe} ~\tilde{g}(x) ~,
 \label{eq.deltai}
\end{equation}
with
\begin{equation}
y_{\rm lobe} = \frac{\sigma_T}{m_{\rm e} c^2}\int
%\varepsilon U_{\rm e} d\ell ~,
P_{\rm e} d\ell ~
 \label{eq.ylobe}
\end{equation}
(see Colafrancesco 2008) where $P_e= \varepsilon U_e$ is the
pressure of electrons
\begin{equation}
P_e=\int_0^\infty dp N(p) \frac{1}{3} p v(p) m_{\rm e} c \, ,
\end{equation}
and $v(p)$ is the velocity corresponding to the momentum $p$; we
note that for a relativistic population of particles $\varepsilon
= 1/3$ and the relation $P_e=U_e/3$ holds.
The relativistic SZE spectral function in eq.(\ref{eq.deltai})
writes
\begin{equation}
 \tilde{g}(x)=\frac{m_{\rm e} c^2}{\langle \varepsilon_{\rm e} \rangle} \left\{ \frac{1}{\tau_e} \left[\int_{-\infty}^{+\infty} i_0(xe^{-s}) P(s) ds-
i_0(x)\right] \right\}
 \label{eq.gdx}
\end{equation}
where $x \equiv h \nu / k T_{CMB}$, in terms of the photon
redistribution function $P(s)$ and of  $i_0(x) = I_0(x)/[2 (k_{\rm
B} T_{CMB})^3 / (h c)^2] = x^3/(e^x -1)$, where
\begin{equation}
 \langle \varepsilon_{\rm e} \rangle  \equiv  \frac{\sigma_{\rm T}}{\tau_e}\int P_e d\ell
%= \int_0^\infty dp f_{\rm e}(p) \frac{1}{3} p v(p) m_{\rm e} c
 \label{temp.media}
\end{equation}
is the average energy of the electron plasma (see Colafrancesco et
al. 2003).
The optical depth of the electron population along the line of
sight $\ell$ passing through the center of the lobe writes
\begin{equation}
 \label{tau_p1}
 \tau_{\rm e}(p_1) = \sigma_T \int d \ell N_{tot}(p_1) = 2\sigma_T R N_{tot}(p_1)
\end{equation}
where $R$ is the radius of the lobe (we assume here a spherical
symmetry and a uniform distribution of the electrons density
within the lobe as in Isobe et al. 2011a), and
$N_{tot}(p_1)=\int_{p_1}^\infty N(p)$.

Since the amplitude and spectral shape of the GRG lobe SZE depends
on the amplitude and spectral shape of the electron spectrum
$N(p)$, the SZE can be used to determine reliably, and in an
unbiased way, the total energy density $U_e$ and to assess the
particle and B-field regime in which RG lobes are found.

\begin{figure}[ht!]
\begin{center}
 \epsfig{file=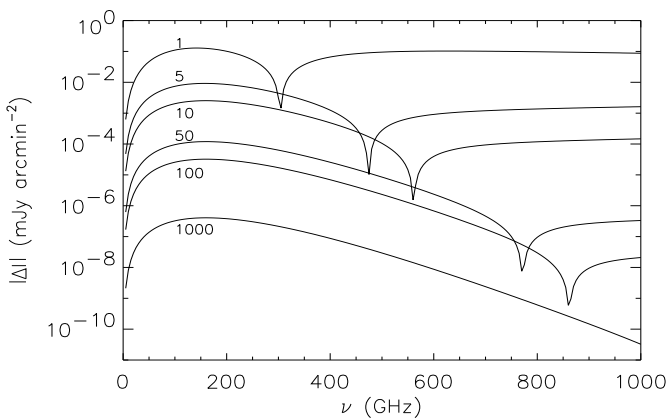,height=6.cm,width=9.cm,angle=0.0}
\end{center}
 \caption{\footnotesize{The absolute value of the SZE spectrum at the center of the
 East lobe of DA 240 for different values of the lowest electron
 momentum $p_1$ (as labelled). We adopt a value of the
 electron spectrum normalization $k_0=1.02 \times 10^{-5}$ cm$^{-3}$ as
 found from the normalization to the Suzaku X-ray spectrum with $p_1=1$ and $\alpha=2.9$.
 }}
 \label{fig.sz_p1}
\end{figure}

The SZE spectrum expected for a population of relativistic
electrons filling the East lobe of DA 240 is shown in
Fig.\ref{fig.sz_p1} and depends sensitively on the value of $p_1$:
in fact, once the values of $\alpha$ and $k_0$ are fixed from
radio and X-ray observations, respectively, both the intensity and
the spectral shape of the SZE (as well as the minimum, crossover
and the maximum of the SZE) depend only on the specific value of
$p_1$.
Fig.\ref{fig.sz_p1} shows the absolute value of $|\Delta
I_{lobe}(x)|$ produced in the E lobe of DA 240 over a wide
frequency range, up to $10^3$ GHz, where the GRG lobe SZE spectrum
spans, and shows its main spectral characteristics: i) the
negative part of the SZE spectrum from low-$\nu$ up to the
crossover frequency (indicated in Fig.\ref{fig.sz_p1} by the dip
in the spectrum located at $\nu \approx 300, 565, 855$ GHz for
$p_1= 1, 10, 100$ respectively) shows a minimum (negative in sign)
of its amplitude (i.e. a maximum in $|\Delta I_{lobe}|$) that is
located at $\nu \approx 150$ GHz independently of $p_1$; ii) a
crossover frequency from negative to positive part of the spectrum
that increases its frequency location for increasing values of
$p_1$; iii) a maximum (positive in sign) of the SZE amplitude
whose frequency location strongly increases with increasing values
of $p_1$.
It is interesting to notice that the SZE for $p_1 = 10^3$ remains
always negative at all frequencies less than $10^3$ GHz, and with
a minimum amplitude of $\Delta I \approx - 3 \cdot 10^{-7}$ mJy
arcmin$^{-2}$ (this is plotted as a maximum in the value of
$|\Delta I_{lobe}|$ in Fig.\ref{fig.sz_p1}). For lower values of
$p_1$, the SZE spectrum shows both the negative and positive parts
of the spectrum. The strong differences of the SZE spectrum allow,
hence, to distinguish various values of $p_1$ from a
multi-frequency SZE analysis.
This strong dependence allows to use observations of the SZE to
measure $p_1$ from a wide-frequency ($\sim 10^2 \div 10^3$ GHz)
analysis of the SZE in GRG lobes.

The non-thermal SZE spectrum of DA 240 shown in
Fig.\ref{fig.sz_p1} has been computed following the approach
described in Colafrancesco (2008) using the spectral index
$\alpha=2.9$ from low-$\nu$ radio data and the normalization $k_0$
derived from the assumption that the X-ray flux at 1 keV from
Suzaku is due to ICS-CMB and has a spectrum consistent with
$\alpha=2.9$.
We note here that this is a valid approach if the value of $p_1$
is sufficiently low in order to ensure a true power-law ICS-CMB
X-ray spectrum for recovering the electron spectrum normalization
$k_0$ from X-ray measurements. Actually for $p_1=10^3$ the effect
of the spectrum cutoff becomes visible at 1 keV and it is
necessary to use a higher normalization of the electron spectrum
to reproduce the X-ray ICS-CMB data (see Sect.2). Observations of
ICS-CMB in the hard X-ray band (with Astro-H or NuSTAR) or in the
gamma-ray band (with Fermi and the future CTA in the high-energy
gamma-rays, and soft gamma-ray spectrometers) would better recover
the power-law spectrum indicated by the low-$\nu$ radio data and
are therefore more suitable for a reliable normalization with high
values of $p_1$.

Once the value of $p_1$ has been derived from the SZE analysis
(see Colafrancesco et al. 2011 for a detailed discussion of a
sample of RG lobes), the total energy density of the electron
population in the RG lobe can be directly obtained using
eq.(\ref{eq.ene.ele}) (see Colafrancesco 2008, Colafrancesco et
al. 2011).
Tab. \ref{tab.1} shows the values of $U_e$ derived for three GRG
lobes observed with Suzaku: the East lobe of DA 240 ($\alpha=2.9$,
$k_0=1.02\times10^{-5}$ cm$^{-3}$), the West lobe of 3C326
($\alpha=2.6$, $k_0=3.24\times10^{-6}$ cm$^{-3}$) and the lobes of
3C35 ($\alpha=2.4$, $k_0=4.91\times10^{-7}$ cm$^{-3}$).
\begin{table}[htb]{}
\vspace{2cm}
\begin{center}
\begin{tabular}{|*{4}{c|}}
\hline $p_1$& $U_e$ (DA 240) & $U_e$ (3C 326) & $U_e$ (3C 35)\\
    & erg cm$^{-3}$ & erg cm$^{-3}$ & erg cm$^{-3}$\\
\hline 1    & $6.15\times10^{-12}$ & $3.21\times10^{-12}$ &
$7.91\times10^{-13}$\\ 5    & $1.98\times10^{-12}$ &
$1.56\times10^{-12}$ & $4.98\times10^{-13}$\\ 10   &
$1.11\times10^{-12}$ & $1.07\times10^{-12}$ &
$3.88\times10^{-13}$\\ 50   & $2.71\times10^{-13}$ &
$4.18\times10^{-13}$ & $2.08\times10^{-13}$\\ 100  &
$1.46\times10^{-13}$ & $2.77\times10^{-13}$ &
$1.58\times10^{-13}$\\ 1000 & $1.85\times10^{-14}$ &
$6.99\times10^{-14}$ & $6.27\times10^{-14}$\\ \hline
\end{tabular}
\end{center}
\caption{\footnotesize{Values of the electron energy density $U_e$
for DA 240 East, 3C 326 West and 3C 35. }}
 \label{tab.1}
\end{table}

\begin{figure}[ht!]
\begin{center}
 \epsfig{file=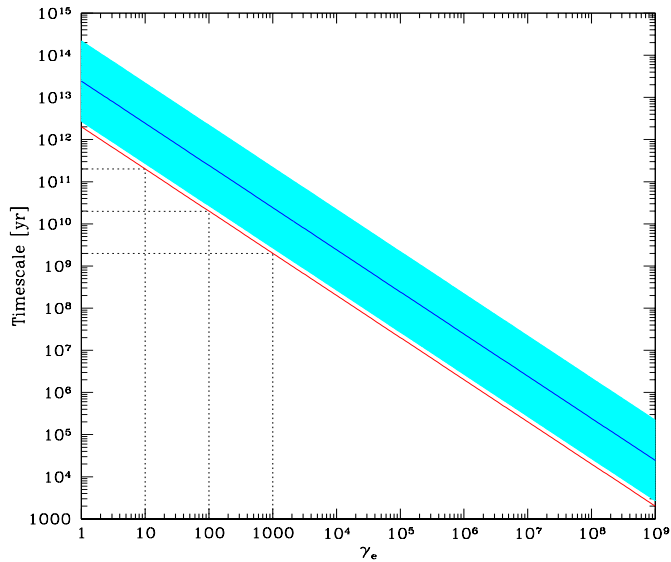,height=8.cm,width=9.cm,angle=0.0}
\end{center}
 \caption{\footnotesize{The ICS (red) and synchrotron (blue) timescales for electrons
 with energy $\gamma_e$. The cyan shaded region encloses
 synchrotron losses for $B_{\mu}$ in the range 0.3--3.
 }}
 \label{fig.timescale}
\end{figure}

The minimum electron momentum $p_1$ can be directly derived from
SZE observations by using eqs.(\ref{eq.ene.ele}--\ref{eq.gdx}),
with also eqs.(21-22) from Colafrancesco (2008) for an analytical
expression for $U_e$, that yield for steep spectrum sources, like
GRG lobes with $\alpha > 2$, the equation:
\begin{equation}
 {1 \over p_1^{1-\alpha}}
  \bigg[{1 \over 2} B_{\frac{1}{1+p_1^2}} \left(\frac{\alpha-2}{2}, \frac{3-\alpha}{2}\right)
   + p_1^{1 -\alpha} \left( (1+p_1^2)^{1/2} -1 \right) \bigg]
   \nonumber
\end{equation}
\begin{equation}
   = {\Delta I_{lobe} \over I_0} {1 \over 2 \varepsilon R \sigma_T k_0
   \tilde{g}(x)} \; ,
 \label{eq.p1sze}
\end{equation}
that can be inverted to obtain $p_1$ from the observable
quantities $\Delta I_{lobe}, R, k_0$ at the frequency $x$.\\
Once the parameter $p_1$ (or equivalently $\gamma_{min}$) is
obtained,  it allows also to estimate the ICS-CMB radiative
timescale of the electron population in GRG lobes that is given by
\begin{equation}
 \tau_{ICS} = {E \over (dE/dt)_{ICS}}
 \label{eq.tauics}
\end{equation}
and is evaluated at the minimum electron energy $\gamma_{min}$.
It turns out that $\tau_{ICS} \approx 2 \cdot 10^8$ yr for
$\gamma_{min}=10^3$, and it is longer for lower values of
$\gamma_{min}$ (see Fig.\ref{fig.timescale}). The radiative
timescale $\tau_{ICS}$ is always shorter than the synchrotron loss
time scale $\tau_{synch}={E \over (dE/dt)_{synch}}$ for all
electron energies responsible for the radio emission in the
326--609 GHz range for values of $B_{\mu}$ fitting the radio
emission, and it is also shorter than the Hubble time $t_H =
H_0^{-1} \approx 9.8 h^{-1} 10^9$ yr for $\gamma \simgt 144$.
Typical spectral ageing and dynamical analysis for GRGs (see,
e.g., Jamrozy et a. 2005, and references therein) show that the
dynamical age of GRG lobes are $\simgt 4$ times the maximum
synchrotron age of the emitting electrons, and hence $\simgt 2
\cdot 10^8$ yr (see the specific case of the GRG J1343+3758
studied by Jamrozy et al. 2005). We notice here that the spectral
ageing technique relies on the electron energy correspondent to
the frequency of the break in the GRG radio spectrum which is
hence a quite high electron energy (of order of several to several
tens GeV) with respect to the minimum electron energies observed
in GRG radio spectra (which are of order of a few GeV) thus
yielding lower limits to the maximum radiative electron timescale.
The ICS-CMB emission provides indications on even lower electron
energies that hence probe larger electron radiative timescales. We
conclude, therefore, that the SZE analysis of GRG lobes can
provide an estimate of the maximum electron radiative timescale
and hence an estimate of the lobe dynamical timescale $t_{dyn}
\approx (1 + \eta) \tau_{rad}$, with $\eta$ taking values of a
few, if the lobe suffers substantial inflation from backflows (see
discussion in Jamrozy et al. 2005).\\
This argument holds for electrons with relatively high energies as
to satisfy $\tau_{ICS} \simlt \tau_H$ and $\tau_{ICS} \simlt
\tau_{Coul}$ or $\tau_{ICS} \simlt \tau_{brem}$ since at low
energies (less than a few hundreds MeV) Coulomb losses and
bremsstrahlung losses dominate the radiative time-scales depending
on the value of the ambient density (we remind the reader that
$\tau_{ICS}= \gamma/[1.37\times10^{-20} \gamma^2 (1+z)^4]$ s,
$\tau_{sync}= \gamma /(1.30\times10^{-21} \gamma^2 B_\mu^2)$ s,
$\tau_{brem}= \gamma/\{1.51\times10^{-16} n_e \gamma [\ln \gamma
+0.36] \}$ s and $\tau_{Coul}= \gamma/ \{1.2\times10^{-12} n_e
\{1.0+[\ln(\gamma/n_e)]/75\}\}$ s, where $n_e$ is the number
density of the ambient thermal electrons; see, e.g.,
Colafrancesco, Profumo \& Ullio 2006, see also Longair 1993,
Sarazin 1999).

The presence of a substantial thermal electron population provides
a possible complication to the previous picture due to the
presence of a fraction of thermal electrons that are heated by the
processes that accelerate the majority of non-thermal electrons.
In such a case, a thermal SZE is produced in addition (or
alternatively if the acceleration mechanism is inefficient) to the
non-thermal relativistic SZE previously discussed (see also
discussion in Yamada et al. 2010).
\begin{figure}[ht!]
\begin{center}
 \epsfig{file=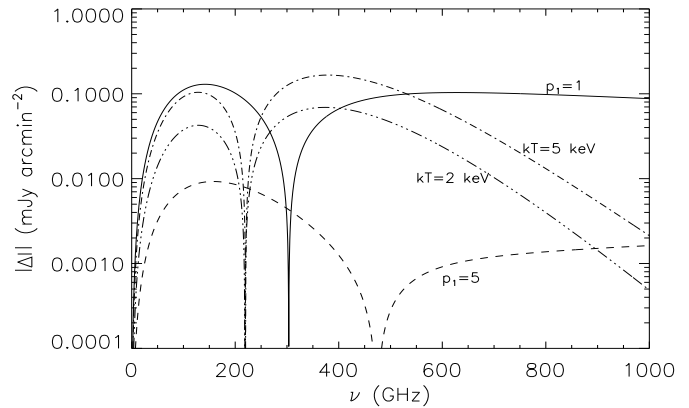,height=6.cm,width=9.cm,angle=0.0}
\end{center}
 \caption{\footnotesize{The non-thermal SZE in a GRG
 lobe like the E lobe of DA 240 is shown for two
 values of $p_1=1,5$ (solid and dashed lines, respectively)
and it is compared with the thermal SZE in the same environment
for two values of the thermal electron
 temperature $kT=2,5$ keV and a reference density $n_e=10^{-4}$ cm$^{-3}$
(three dots-dashed and dot-dashed lines, respectively).
 We note that the lines corresponding to the non-thermal
 SZE are the same that are plotted in Fig.\ref{fig.sz_p1}.
}}
 \label{fig.sz_th+nt}
\end{figure}
To discuss this possible option, we show in Fig.\ref{fig.sz_th+nt}
the case in which a thermal and a non-thermal SZE from the E lobe
of DA 240 are generated. The non-thermal SZE is shown for the
electron spectrum as in Fig.\ref{fig.sz_p1} and for the two values
of $p_1=1$ and $5$. The possible thermal SZE in the lobe has been
computed for two values of the thermal electron temperature of
$kT=2$ and $5$ keV and assuming a reference density $n_e=10^{-4}$
cm$^{-3}$ of the thermal particles in the lobe.\\
Again a multi-frequency analysis of the SZE signals (but with a
higher precision of the observations, see Colafrancesco \&
Marchegiani 2010 for a quantitative discussion) over a large
frequency range is able to disentangle the two sources of the SZE
and hence the two electron populations.

\section{Other RG lobes}

The GRG DA 240 is only a representative case of an increasing
database of GRGs with lobes studied through ICS-CMB emission.
Studies of the energetics of GRG lobes through X-ray ICS-CMB
observations become statistically relevant with the advent of
Chandra, XMM and Suzaku.
To widen our discussion, we compare in this Section our findings
for DA 240 with other two GRGs observed with Suzaku (see
Tab.\ref{tab.1}) and with the sample of FR II RG lobes studied in
X-rays by Croston et al. (2005). For all these objects,
information on $U_e$ and $U_B$ are available from the combination
of radio and X-ray data.

\begin{figure}[ht!]
\begin{center}
 \epsfig{file=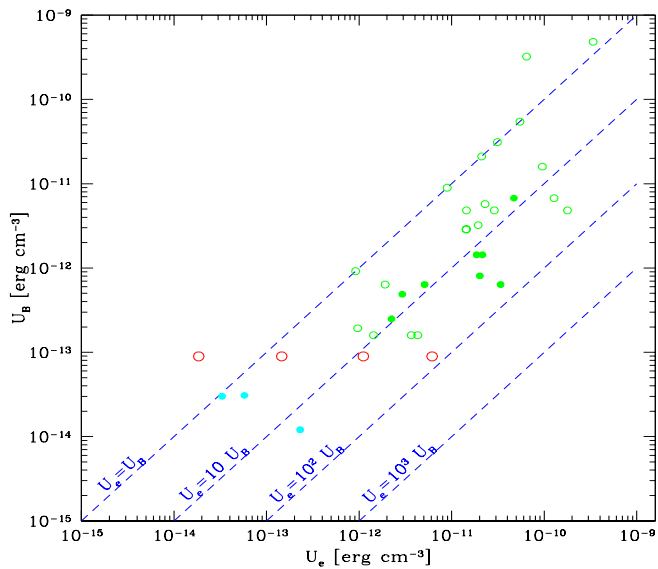,height=8.cm,width=9.cm,angle=0.0}
\end{center}
 \caption{\footnotesize{The trajectory of the RG lobe DA240 in the $U_e - U_B$
 plane calculated with $B_{\mu}=1.5$ (that fits the radio spectrum with $p_1=1$)
 for decreasing values of $p_1=10^3-1$ (from left to right) is shown by the red open circles.
 The results of Isobe et al. (2009, 2011a,b) for the GRGs DA240E, 3C35 and 3C326W
 assuming $p_1=10^3$ are shown by the cyan dots. RG lobes from the analysis of Croston
 et al. (2005) obtained assuming $\alpha=2$ and $p_1=10$ for all lobes
 are also shown as green dots: filled dots refer to lobes with
 reliable X-ray spectral estimates. Note these last data are not directly
 comparable to those of Isobe et al. (2009, 20011a, 2011b) for GRGs.
 }}
 \label{fig.ueub}
\end{figure}

Fig.\ref{fig.ueub} shows the position of the three GRG lobes DA
240E, 3C326W and 3C35 in the $U_e - U_B$ plane.
We find that GRG lobes are likely away from equipartition for any
value $p_1 < 10^3$ and that their energy density is largely
particle dominated.
The RG lobes analyzed by Croston et al. (2005), under the
assumption $p_1=10$ and a constant power-law index $\alpha=2$ for
all the RG lobes, also indicate $U_e/U_B \sim 10 - 10^2$ when
accurate X-ray spectra are obtained (filled green dots in
Fig.\ref{fig.ueub}). The other RG lobes with an X-ray detection
but without reliable spectra (a necessary information to fully
characterize the ICS-CMB emission) are subject to strong
uncertainties in determining the nature of their X-ray emission
(see, e.g., discussion in  Croston et al. 2005). In fact, their
distribution in the $U_B-U_e$ plane is very sparse and uncertain,
with the lobes distributed over very different physical regimes,
from below-equipartition condition ($U_e < U_B$) to a regime of
strong particle dominance ($U_e \simgt (10-100) U_B$).\\
The shape of the correlation between $U_e$ and $ U_B$ for RG lobes
depends, as we have already discussed, on the specific value of
$p_1$ for each RG lobe, and it can show if GRGs are systematically
different from normal RGs in their energetics and therefore in
their lower energy spectrum cutoff.

\begin{figure}[ht!]
\begin{center}
 \epsfig{file=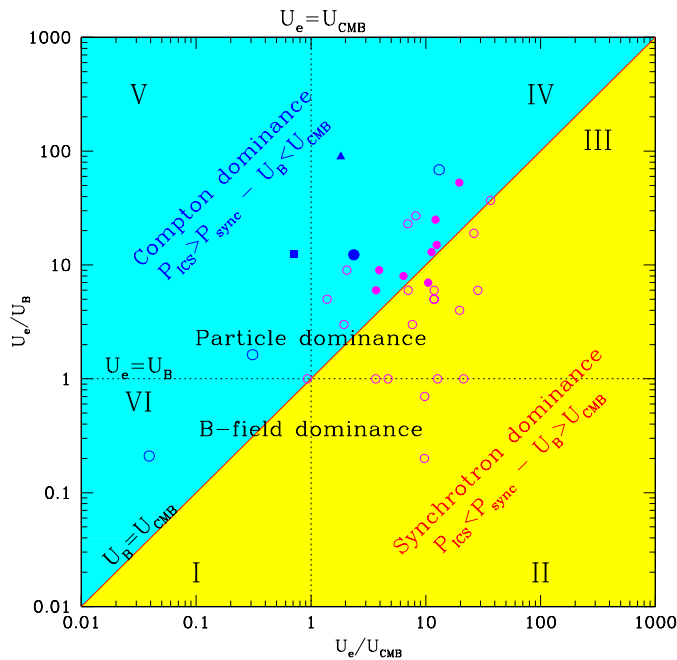,height=9.cm,width=9.cm,angle=0.0}
\end{center}
 \caption{\footnotesize{The trajectory of the GRG lobes of DA240 (open blue circles),
 in the $U_e/U_B$ vs. $U_e/U_{CMB}$ plane is shown for decreasing values of
 $p_1$ from $10^3$ to $1$ as in Tab.\ref{tab.1}. The positions of
 DA 240 (filled blue circle), 3C35 (blue filled square) and 3C326
 (blue filled triangles) for $p_1=10$ are also shown.
 The shaded areas show the regions of Compton (radiation) dominance
 ($P_{ICS}>P_{synch}$ or $U_B < U_{CMB}$: cyan) and synchrotron (B-field) dominance
 ($P_{ICS}<P_{synch}$ or $U_B > U_{CMB}$: yellow) for the single particle limit. The
 equipartition lines $U_e=U_B$, $U_e=U_{CMB}$ and $U_B = U_{CMB}$ are also shown
 for comparison. The magenta dots show the RG lobes analyzed by Croston et al. (2005)
 assuming $p_1=10$ and $\alpha = 2$:
 filled magenta dots refer to lobes with reliable X-ray spectral estimates
 (see text for discussion).
 }}
 \label{fig.uratio}
\end{figure}

To further discuss the point, we also show in Fig.\ref{fig.uratio}
the trajectory of DA 240 in the $U_e/U_B$ vs. $U_e/U_{CMB}$ plane
for values $p_1=1, 10, 10^2, 10^3$ and the location of 3C35 and
3C326 for $p_1=10$ in order to compare these objects with the RG
lobes analyzed by Croston et al. (2005) under the same assumptions
(i.e. $p_1=10$).

The plot of Fig.\ref{fig.uratio} is useful to determine the
physical conditions of each RG lobe. The different sectors in this
plane (see Tab. \ref{tab.2}) specify if a RG lobe is particle or
radiation dominated, Compton or synchrotron dominated and in which
equipartition regime among $U_e, U_B, U_{CMB}$ they are found.
Tab.\ref{tab.2} describes synthetically the different physical
regimes for the considered lobes in the $U_e/U_B$ vs.
$U_e/U_{CMB}$ plane.
\begin{table}[htb]{}
\vspace{2cm}
\begin{center}
\begin{tabular}{|*{4}{c|}}
 \hline
Region & Regime                 & Condition           & Notes \\
 \hline
 I     & Synch. dominance       & $P_{ICS}<P_{synch}$ & Synch. \\
       & B-field dominance      & $U_{B}>U_{CMB}$     & High-B        \\
       &                        & $U_{e}<U_{B}$       & Low $N(p)$    \\
       &                        & $U_{e}<U_{CMB}$     &               \\
 \hline
 II    & Synch. dominance       & $P_{ICS}<P_{synch}$ & Synch. \\
       & B-field dominance      & $U_{B}>U_{CMB}$     & High-B        \\
       &                        & $U_{e}<U_{B}$       & High $N(p)$    \\
       &                        & $U_{e}>U_{CMB}$     &               \\
 \hline
 III   & Synch. dominance       & $P_{ICS}<P_{synch}$ & Synch. \\
       & Particle dominance  & $U_{B}>U_{CMB}$        & Low-B        \\
       &                        & $U_{e}>U_{B}$       & High $N(p)$    \\
       &                        & $U_{e}>U_{CMB}$     &               \\
 \hline
 IV    & Compton dominance      & $P_{ICS}>P_{synch}$ & ICS-CMB \\
       & Particle dominance     & $U_{B}<U_{CMB}$     & Low-B        \\
       &                        & $U_{e}>U_{B}$       & High $N(p)$    \\
       &                        & $U_{e}>U_{CMB}$     &               \\
 \hline
 V     & Compton dominance      & $P_{ICS}>P_{synch}$ & ICS-CMB  \\
       & Particle dominance     & $U_{B}<U_{CMB}$     & Low-B        \\
       &                        & $U_{e}>U_{B}$       & Low $N(p)$    \\
       &                        & $U_{e}<U_{CMB}$     &               \\
 \hline
 VI    & Compton dominance      & $P_{ICS}>P_{synch}$ & ICS-CMB \\
       & B-field dominance      & $U_{B}<U_{CMB}$     & High-B        \\
       &                        & $U_{e}<U_{B}$       & Low $N(p)$    \\
       &                        & $U_{e}<U_{CMB}$     &               \\
\hline
\end{tabular}
\end{center}
\caption{\footnotesize{Physical regimes for RG lobes in the
$U_e/U_B$ vs- $U_e/U_{CMB}$ plane of Fig.\ref{fig.uratio}.}}
 \label{tab.2}
\end{table}

The GRG lobes of DA 240, 3C326 and 3C35 seem to be completely
particle dominated and in the Compton power dominance regime in
which their total power is dominated by the ICS-CMB emission ($U_B
< U_{CMB}$ and $(dE/dt)_{ICS} > (dE/dt)_{syn}$ ) off the electron
population residing in their lobe.

To demonstrate the predictive power of our analysis, we notice
that all of the Croston et al. (2005) RG lobes with good X-ray
spectra are also particle-dominated and Compton power dominated
except one, i.e. 3C265W: this last lobe has, in fact, the highest
value of the magnetic field $B_{\mu}=13$. It has also been
suggested that the ICS nuclear scattering emission could
contribute substantially to its X-ray flux (see, e.g., Bondi et
al. 2004, Croston et al. 2005), thus yielding larger uncertainties
on the nature of the X-ray emission and possibly overestimating
artificially the value of $U_e/U_{CMB}$.
If we however take the value of $U_e/U_B$ for 3C265W at face
value, we should conclude that it is both particle dominated ($U_e
> U_B$) and synchrotron dominated $U_B > U_{CMB}$ due to the
rather high value of $B_{\mu}=13$, which do not make this lobe a
clean representative of ICS-CMB dominated system.

The extreme case of the 3C280 lobe is found to be B-field and
synchrotron dominated (lower-right quadrant II in
Fig.\ref{fig.uratio}) and it has the lowest value of
$U_e/U_B=0.2$.
This RG lobe suffers, however, from a lack of definite spectral
measurement at X-rays that does not allow to determine the nature
of its X-ray emission. The difficulty of separating correctly the
different components of the X-ray emission in RG lobes led to
discrepant results in the literature (Croston et al. 2005). In
this respect, it is extremely unlikely that any flux measurements
are under-estimates of the lobe ICS-CMB emission, so that any
systematic uncertainty in the value of $U_e/U_B$ and $U_e/U_{CMB}$
is likely to be in the direction of overestimation.\\
Our analysis of RG lobes reported in Fig.\ref{fig.uratio}
indicates that all GRG lobes with definite ICS-CMB spectral
measurements are particle-dominated and ICS-CMB power dominated.
The other RG lobes have X-ray spectral measurements likely
contaminated by nuclear or hot-spot X-ray emission.

We propose here to use the $U_e/U_B - U_e/U_{CMB}$ plot to better
characterize the physical regime of RG lobes, and we will present
a more extensive analysis of RG lobes multi-frequency emission
elsewhere (Colafrancesco et al. in preparation).

The condition $U_e/U_B \sim 10-100$ suggested by the GRG lobe data
(see Figs.\ref{fig.ueub} and \ref{fig.uratio}) indicates the
presence of a substantial amount of electrons with energies above
$\gamma_{min} \simgt 10$ and up to $\gamma_{max} \sim 10^{5}-10^6$
as seems to be required in order to explain the radio, mm, X-ray
and (in a few observed cases) also the gamma-ray observations
(see, e.g., Massaro \& Ajello 2011 and references therein). Such
relatively energetic electron population spread over the GRG lobes
could be sustained by turbulent acceleration mechanisms and
diffusion in a filamentary B-field configuration (see
Gopal-Krishna et al. 2001) that seem to be efficient in balancing
their radiative energy losses along the late stage of the GRG lobe
evolution.

More effective and definitive observations are required to probe
the total energy density of RG lobes however. The results reported
here demonstrate that SZE measurements are crucial to establish in
an unbiased way the realistic estimates of $U_e$ and hence the
correlation between $U_e$ and $U_B$ for RG lobes.

\section{Summary and conclusions}

We have shown that the electron energy density $U_e$ in GRG lobes
derived from X-ray observations is only a rough lower limit to the
actual value of $U_e$ if $p_1$ is substantially less than $10^3$,
as indicated by observations and theoretical expectations. This
result is even stronger for larger spectral indices $\alpha$, as
those shown by GRG lobes.
This indicates that -- even though X-ray measurements are relevant
to set the electron spectrum normalization at high-$E_e$ -- the
properties of the low-energy electron population in the radio
lobes are not well constrained by X-ray measurements that probe
electrons with $E_e \sim 0.1 - 1$ GeV (for $E_X \sim 0.1-10$ keV),
a particularly problematic issue when studying the overall power
of the ICS-CMB process. This further leads to reconsider the
physical regimes in which GRG lobes are found.\\
We have shown that all RG lobes with reliable ICS-CMB spectral
measurements in the X-ray band are in a regime of radiative
Compton dominance and particle domination with respect to the
B-field energy density. Therefore RG lobes are likely expected to
have $U_e/U_B \simgt 10-100$ and $U_e/U_{CMB} \simgt 1-20$ which
indicate that they have a population of relatively energetic
electrons providing a substantial energy and pressure support to
the lobes, consistent with a turbulent MHD acceleration origin.
Under these regime, we predict that many other RG lobes will be
visible with the next generation hard X-ray (Astro-H, NuSTAR) and
some also with gamma-ray (CTA and the next-generation soft
gamma-ray spectrometer) instruments. The importance of the future
hard X-ray and gamma-ray observation is their ability to set the
electron spectrum normalization, even though they cannot
definitely assess the value of the total $U_e$. We have shown that
the SZE visible in the direction on many of these GRG lobes with
the next coming microwave and mm spectroscopic experiments (see
discussion in Colafrancesco \& Marchegiani 2010 and Colafrancesco
et al. 2011) ensures the capability of determinig the total value
of $U_e$ since the SZE is sensitive to the overall energy spectrum
extension and in particular to the minimum energies of the
electron population.

A final point we stress is that the determination of the RG lobe
B-field from radio and X-ray observations implies the assumption
of a model for the electron spectrum (as also pointed out in
Croston et al. 2005). In fact, electrons that emit in the radio
(by synchrotron) and in the X-rays (by ICS-CMB) are not the same
ones except for a specific choice of the B-field.
For the case of DA 240 East lobe, an X-ray observation at 1 keV
implies that these electrons have an energy of $E_e\sim0.35$ GeV,
and that the same electrons emit synchrotron at a frequency
$\nu\sim2 B_\mu$ MHz with $B_{\mu}=1.5 \div 0.87$ for
$p_1=1-10^3$, respectively.
Electrons that emit synchrotron in the observed range
($\nu=326\div609$ MHz) have energies $E_e\sim(4.5\div6.2)
B_\mu^{-1/2}$ GeV, and emit by ICS-CMB in the range $E_X \sim(
160\div310) B_\mu^{-1}$ keV.
The assumption of using the same electron spectrum in the two
electron energy ranges is likely reasonable only if the radio and
X-ray spectral indices are compatible within the uncertainties.

Future studies of RG lobes with high-sensitivity radio experiments
(like SKA and its precursors, MeerKAT in South Africa and ASKAP in
Australia) can measure the electron spectrum and the magnetic
field with high precision and spatial resolution from observation
over a wide frequency range $\sim 0.1 - 45$ GHz, and their
high-$\nu$ bands at $\simgt 30$ GHz can also be able to measure
simultaneously the radio and SZE spectrum at low frequencies,
while mm experiments (e.g. Planck, OLIMPO, HERSCHEL, ALMA) will be
able to fully measure the SZE spectrum and the RG lobe energetics.
In this context, hard X-rays (NuSTAR, Astro-H) and gamma-rays
(Fermi, CTA) will be crucial to determine the high-E spectrum
normalization and to have indication on the high-E energy cutoff.
The recent observations on the Centaurus A with the Fermi-LAT
instrument (Abdo et al. 2010) and of Fornax-A with Suzaku (Tashiro
et al. 2009) indicate that ICS-CMB emission dominates at very high
frequencies the GRG lobe SEDs and open the way to the
determination of their electron spectrum at the high-E tail. These
observations will be crucial for SZE observational techniques of
GRG lobes to be fully exploited.

\begin{acknowledgements}
We thank an anonymous Referee for several useful comments and
suggestions that allowed to improve the discussion and the
presentation of our results. S.C. acknowledges support by the
South African Research Chairs Initiative of the Department of
Science and Technology and National Research Foundation and by the
Square Kilometre Array (SKA).
\end{acknowledgements}

%%%%%%%%%%%%%%%%%%%%%%%%%%%%%%%%%%%%%%%%%%%%%

%%%%%%%%%%%%%%%%%%%%%%%%%%%%%%%%%%%%%%%%%%%%%%%%%%%%%%%%%%%%%%%%%%%%%%%

\end{document}